\documentclass[10pt,conference]{IEEEtran}
\ifCLASSOPTIONcompsoc
\usepackage[nocompress]{cite}
\else
\usepackage{cite}
\fi
\ifCLASSINFOpdf
\else
\fi
\usepackage{authblk}
\usepackage{mathtools}
\DeclarePairedDelimiter\floor{\lfloor}{\rfloor}

\usepackage{amsmath,amssymb,amsfonts}
\usepackage{algorithm}
\usepackage{algpseudocode}
\usepackage{mdframed}
\usepackage{graphicx}
\usepackage{subfig}
\usepackage{multicol}
\usepackage{textcomp}
\usepackage{xcolor}

\def\BibTeX{{\rm B\kern-.05em{\sc i\kern-.025em b}\kern-.08em
		T\kern-.1667em\lower.7ex\hbox{E}\kern-.125emX}}
\newcommand{\comment}[1]{}

\newtheorem{theorem}{Theorem}

\newtheorem{problem}{Problem}
\newtheorem{definition}{Definition}
\newtheorem{proposition}{Proposition}

\usepackage{mdframed}

\begin{document}
	%
	\title{Optimization of ARQ Distribution for HARQ Strategies in Delay-Bounded Networks}
	
	%
	%
	%
	%
	
	\author{Jaya Goel and J. Harshan\\
	Bharti School of Telecom Technology and Management,\\
		Indian Institute of Technology Delhi, India
	}
	
	
	%
	%
	
	\markboth{Journal of \LaTeX\ Class Files,~Vol.~XX, No.~X, XXXX}%
	{Shell \MakeLowercase{\textit{et al.}}: Bare Demo of IEEEtran.cls for Computer Society Journals}
	%



	\IEEEtitleabstractindextext{%
	\begin{abstract}
Inspired by several delay-bounded mission-critical applications, optimizing the end-to-end reliability of multi-hop networks is an important problem subject to end-to-end delay constraints on the packets. Towards that direction, Automatic Repeat Request (ARQ) based strategies have been recently proposed wherein the problem statement is to distribute a certain total number of ARQs (that capture end-to-end delay) across the nodes such that the end-to-end reliability is optimized. Although such strategies provide a fine control to trade end-to-end delay with end-to-end reliability, their performance degrades in slowly-varying channel conditions. Pointing at this drawback, in this work, we propose a Chase Combing Hybrid ARQ (CC-HARQ) based multi-hop network addressing a similar problem statement of how to distribute a certain total number of ARQs such that the end-to-end reliability is optimized. Towards solving the problem, first, we identify that the objective function of the optimization problem is intractable due to the presence of Marcum-Q functions in it. As a result, we propose an approximation on the objective function and then prove a set of necessary and sufficient conditions on the near-optimal ARQ distribution. Subsequently, we propose a low-complexity algorithm to solve the problem for any network size. We show that CC-HARQ based strategies are particularly appealing in slow-fading channels wherein the existing ARQ strategies fail.  
		\end{abstract} 
		
		}

	\maketitle

	\IEEEdisplaynontitleabstractindextext

	%
	\IEEEpeerreviewmaketitle
	
\section{Introduction}
\label{sec:intro}

In the fifth generation (5G) wireless networks and beyond, Internet-of-Things (IoTs), massive Machine-type communications (mMTC) etc., \cite{Schulz2017} are pitched to play a vital role in deploying mission-critical applications. Examples include reliable and timely delivery of status updates in vehicular networks, facilitating real-time industrial automation tasks by deploying massive low-power IoT devices in factory settings, and so on. Be it vehicular networks or industrial IoT settings, one of the main challenges is the problem of designing signalling schemes over large-scale devices and understanding their fundamental limits in achieving ultra-reliable communication with bounded delay constraints on the packets. Furthermore, generalizing these mission-critical problem statements to multi-hop wireless network settings is also important owing to its proven expansion in the coverage area \cite{Badarneh2016}, especially when involving low-powered devices. Thus, in this work, we address the challenges in envisioning delay-bounded mission-critical applications over multi-hop wireless networks \cite{Shaikh2018, Ji2018}.   

Ongoing research in wireless multi-hop networks has already shown the challenges in achieving high end-to-end reliability with bounded delay-constraint on the packets. \cite{wiopt, our_work_TWC_1} recently proposed a decode-and-forward (DF) based relaying strategy along with retransmission protocols at the intermediate nodes. In such a scheme, reliability within a hop was taken care of by the number of retransmissions allotted to a given node, and the end-to-end delay on the packets was dictated by the total number of ARQs allotted to all the nodes in the multi-hop network. By using the information on processing delays at each node, a certain total number of retransmissions was estimated corresponding to the deadline on the packets. 

In order to understand the relation between the bounded delay constraints and the total number of retransmissions, we first explain the \emph{worst-case deadline} approach of computing the maximum number of retransmissions by assuming that the delay overheads from ACK/NACK in the reverse channel are sufficiently small compared to the payload. Suppose that the processing time at each hop is $\tau_{p}$ seconds (which includes packet encoding and decoding time), the delay incurred for packet transmission at each hop is $\tau_{d}$ seconds (which includes the propagation delay and the time-frame of the packet), and the delay incurred because of NACK overhead is $\tau_{NACK}$ (which is the time taken for the transmitter to receive the NACK). Given the stochastic nature of the wireless channel at each link, the total number of packet retransmissions before the packet reaches the destination is a random variable, denoted by $n$, and as a result, the end-to-end delay between the source and the destination is upper bounded by $n \times(\tau_{p} + \tau_{d} + \tau_{NACK})$ seconds. In particular, when $\tau_{NACK} << \tau_{p} + \tau_{d}$, the end-to-end delay can be approximated as $n \times(\tau_{p} + \tau_{d})$ seconds. Thus, when the packet size and the decoding protocol at each node are established, and when the deadline on end-to-end delay (denoted by $\tau_{total})$ is known, we may impose an upper bound on $n$, provided by $q_{sum} = \floor{\frac{\tau_{total}}{\tau_{p}+\tau_{d}}}$. The values of $\tau_{p}$ and $\tau_{d}$ at each relay node are assumed to be identical in order to obtain a relation between the end-to-end delay $\tau_{total}$ and the total number of ARQs $q_{sum}$. While identical $\tau_{d}$ holds in practice as the packet length is the same at each hop, the propagation delay is usually negligible. However, if $\tau_{p}$ at each relay are not identical, then an upper bound on the total number of ARQs can still be obtained by considering the maximum of the processing delays offered by all the relays in the chain. Overall, this approach implies that $q_{sum}$ captures the maximum number of retransmissions that can be tolerated over the multi-hop network in order to respect the deadline on the delay. 

Once $q_{sum}$ is obtained for a given bounded-delay constraint, several questions were posed in \cite{our_work_TWC_1} on how to distribute these retransmissions across different nodes to maximize end-to-end reliability. In particular, Type-1 Automatic Repeat Request (ARQ) protocol was employed, wherein the receiver node at each hop transmits an ACK or NACK depending on the success in decoding. Subsequently, on every retransmission, the receiver node discards the previous version of the packet and only uses the latest packet to decode the information. We note that when these solutions are used in the scenario where the devices are static and their surrounding does not change over time, e.g., factory settings, Type-1 ARQs are not applicable for two apparent reasons: Firstly, in Type-1 ARQ, the previous packet is discarded; therefore, it implies the wastage of resources, and secondly, as the channel may not vary with time, there is no benefit in discarding the previous packets and then decoding the new packet that is re-transmitted under the same channel conditions. In such cases, advanced ARQ strategies, such as Type-2 ARQs \cite{Shen_2011} is beneficial, wherein on every retransmission, the receiver node combines the latest packet along with its previous copies to decode the information. Although an optimization problem on ARQ distribution has been studied in \cite{wiopt, our_work_TWC_1} for Type-1 ARQs, there is no such effort on Type-2 ARQs to achieve high reliability in delay-bounded scenarios. Among many variants of Type-2 ARQs schemes that are often used in practical systems, a popular scheme is Chase Combining Hybrid ARQ (CC-HARQ), wherein each retransmission block is identical to the original code block \cite{Chaitanya_2013}.

Motivated by the drawbacks of using Type-1 ARQs in wireless networks with slowly varying channels, in this work, we propose a CC-HARQ based DF strategy for achieving high reliability under delay-bounded scenarios. In particular, we consider a multi-hop network dominated by slowly varying wireless channels with arbitrary line-of-sight (LOS) components. Along the similar lines of \cite{wiopt}, we impose an upper bound on the total number of ARQs required in the network following the CC-HARQ strategy based on the end-to-end delay requirements of the application. Subsequently, we formulate an optimization problem of allocating the optimal ARQ distribution to each intermediate link such that the packets reach the destination with high reliability.\footnote{Although it appears that allocating $q_{sum}$ ARQs to the first node of the network seems optimal, such an approach would require each node to explicitly communicate the residual ARQs in the packet. This, in turn, leads to additional communication-overhead in the packet, and such a provision may not be allowed in certain applications.} Due to the intractability of the objective function in the optimization problem, first, we propose an approximation for the objective function at a high signal-to-noise-ratio (SNR). After that, we obtain the necessary and sufficient conditions on the near-optimal ARQ distribution followed by a low-complexity algorithm to solve the optimization problem. We show that our CC-HARQ based framework outperforms the Type-1 ARQ based strategy in \cite{wiopt} when channels are slowly varying.

In terms of novelty, this is the first CC-HARQ based strategy that optimizes the reliability aspects of multi-hop networks with bounded-delay constraints. Moreover, the analytical results of this work cannot be viewed as a straightforward extension of prior works since the objective function used for optimization is unique to the CC-HARQ strategy, which was not dealt with bounded-delay applications hitherto.   

For visibility purposes, this manuscript is also made available on arXiv \cite{our_work_arXiv} with the same title. Therefore, \cite{our_work_arXiv} should not be treated as prior-art when evaluating the novelty of this submission.


\section{CC-HARQ Based Multi-Hop Network Model}

\begin{figure}[h!]
\centering \includegraphics[scale = 0.9]{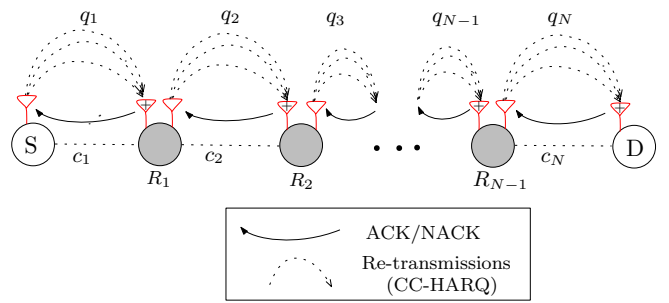}\centering{\caption{Illustration of an $N$-hop network with a source node $(S)$, the relay nodes $R_{1}, \ldots, R_{N-1}$ and the destination node $(D)$ follow the CC-HARQ protocol at each intermediate link. Also, each intermediate link can be characterized by an LOS component $c_{k} \  \forall k \in [N]$. 
\label{system_model}}}
\end{figure}

Consider a network with $N$ hops, as shown in Fig. \ref{system_model}, consisting of a source node (S), a set of $N-1$ relays $R_{1}, R_{2}, \ldots, R_{N-1}$ and a destination node (D). By aggregating the information bits in the form of packets, we communicate these packets from S to D by using the $N-1$ intermediate relays. We assume that the channel between any two successive nodes is characterized by Rician fading that varies slowly over time. We model the complex baseband channel of the $k$-th link, for $1 \leq k \leq N$, as
\begin{equation*}
h_{k} = \sqrt{\frac{c_{k}}{2}}(1+\iota)+\sqrt{\frac{(1-c_{k})}{2}}g_{k},
\end{equation*}
where $\iota = \sqrt{-1}$, $0 \leq c_{k} \leq 1$ is the LOS component, $1-c_{k}$ is the Non-LOS component, and $g_{k}$ is a Gaussian random variable with distribution $\mathcal{CN}(0,1)$. In this channel model, $c_{k}$ is a deterministic quantity, which characterizes different degrees of Rician fading channels, and also makes sure that $\mathbb{E}[|h_{k}|^{2}] = 1$ holds for any $c_{k}$. At the extreme ends, it is well known that $c_{k} = 0$ gives us the Rayleigh fading channels and $c_{k} = 1$ provides the Gaussian channels. 

By assuming that the intermediate relays are sufficiently far apart from each other, throughout the paper, we use the vector $\mathbf{c}= [c_{1},c_{2},\ldots, c_{N}]$ to denote the LOS components of the $N$-hop network.
Let $\mathcal{C} \subset \mathbb{C}^{L}$ denote the channel code used at the source of rate $R$ bits per channel use, i.e., $$R = \frac{1}{L}\mbox{log}(|\mathcal{C}|).$$ Let $\mathbf{x} \in \mathcal{C}$ denote a packet (which is a codeword in the code) transmitted by the source node such that its average energy per channel use is unity. When $\mathbf{x}$ is sent over the $k$-th link, for $1\leq k \leq N$, the corresponding baseband symbols gathered at the receiver over $L$ channel uses is $\mathbf{y}_{k} = h_{k}\mathbf{x} + \mathbf{w}_{k} \in \mathbb{C}^{L}$, where $\mathbf{w}_{k}$ is the additive white Gaussian noise (AWGN) at the receiver of the $k$-th link, distributed as $\mathcal{CN}(0,\sigma^{2}\mathbf{I}_{L})$. We assume that $h_{k}$ is known at the receiver of the $k$-th link owing to channel estimation; however, the transmitter of the $k$-th link does not know $h_{k}$. Since $h_{k}$ is sampled from an underlying distribution and the realization remains constant for $L$ channel uses, the transmission rate $R$ may not be less than the instantaneous mutual information of the $k$-th link. Therefore, in such cases, the corresponding relay node will fail to correctly decode the packet. The probability of such an event is well captured by
	\begin{eqnarray}
		\label{eq:outage_prob_link}
		P_{k} = \mbox{Pr} \Big( R > \log_{2}(1+ |h_{k}|^{2}\gamma ) \Big) = \mbox{F}_{k}\left(\frac{2^{R}-1}{\gamma}\right),
	\end{eqnarray}
	where $\gamma =  \frac{1}{\sigma^{2}}$ is the average signal-to-noise-ratio (SNR) of the $k$-th link, $\mbox{F}_{k}(x)$ is the cumulative distribution function of $|h_{k}|^{2}$, defined as 
	\begin{equation}
		\label{eq:marcum}
		\mbox{F}_{k}\left(\frac{2^{R}-1}{\gamma}\right) = 1-Q_{1}\Bigg(\sqrt{\frac{2c_{k}}{(1-c_{k})}}, \sqrt{\frac{2(2^{R}-1)}{\gamma(1-c_{k})}}\Bigg),
	\end{equation} 
	such that $Q_{1}(\cdot,\cdot) $ is the first-order Marcum-Q function \cite{Marcum} \footnote{It is well known that the expression for $P_{k}$ given in \eqref{eq:outage_prob_link} captures the error expression in asymptotic block-length regimes. However, it has been shown in \cite[Theorem 2]{V_Poor} that \eqref{eq:outage_prob_link} serves as a saddle-point approximation on the error-probability expressions in non-asymptotic block lengths, especially when the block-length $K$ is in the order of a few hundreds.}.
	To support the transmission rate, we follow the CC-HARQ strategy wherein a receiver node asks the transmitter node for retransmission of the packet and combines the received packet with the previous failed attempts to recover the information. To explain further, for a given $N$-hop network model, a transmitter node gets an ACK or NACK from the next node in the chain, indicating the success or failure of the transmission, respectively. Upon receiving a NACK, the transmitter retransmits the packet, and the receiver combines the current packet with previously received copies of the packet. Let $q_{k}$ be the maximum number of attempts given to the transmitter of the $k$-th link to retransmit the packet on demand. Consolidating the number of attempts given to each link, the ARQ distribution of the multi-hop network is represented by the vector $\mathbf{q}= [q_{1},q_{2},\ldots,q_{N}]$. Since we are addressing bounded delay applications, we impose the sum constraint $\sum_{i=1}^{N} q_{i} = q_{sum}$, for some $q_{sum} \in \mathbb{Z}_{+}$, which captures an upper bound on the end-to-end delay on the packets. 
	
	Note that the packet does not reach the destination if an intermediate node fails to deliver the packet to its next node despite using the allotted number of attempts. Since the packet can be dropped in any of the links, we use packet-drop-probability (PDP) as our reliability metric of interest, given by      
	\begin{equation}
		\label{eq:pdp_expression}
		p_{d} = P_{1q_{1}} + \sum_{k=2}^{N} P_{kq_{k}} \Bigg( \prod_{j=1}^{k-1}  (1-P_{jq_{j}})\Bigg),
	\end{equation}
	where $P_{kq_{k}}$ represents the outage event at $k$-th link after using the $q_{k}$ attempts with the CC-HARQ protocol. In particular, we have $P_{kq_{k}}= \mbox{Pr} \left(R > \log_{2}(1+ (\sum_{j=1}^{q_{k}}|h_{kj}|^{2}) \gamma)\right)$, where $h_{kj}$ is the channel realization at the $k$-th link for the $j$-th attempt. The expression on $P_{kq_{k}}$ is obtained due to the maximum ratio combining technique. Under the scenario, when the channel remains fixed over multiple attempts, we can rewrite the above equation as $\mbox{Pr} \left(R > \log_{2}(1+ |h_{k1}|^{2}q_{k} \gamma)\right)$ which is equal to $\mbox{F}_{k}\left(\frac{2^{R}-1}{q_{k}\gamma}\right)$. It can be observed that on every retransmission, the packet gets added to its previous copies due to CC-HARQ protocol, which results in an increased effective SNR of the given link.

For the proposed CC-HARQ based multi-hop network, we are interested in minimizing the PDP given in expression \eqref{eq:pdp_expression} for a given sum constraint on the total number of ARQs given by $q_{sum} \in \mathbb{Z}_{+}$ such that $\sum_{i = 1}^{N} q_{i} = q_{sum}$. Henceforth, we formulate our optimization problem in Problem \ref{opt_problem} as shown below. Throughout this paper, we refer to the solution of Problem \ref{opt_problem}, as the optimal ARQ distribution. 

\vspace{0.5cm}
	\begin{mdframed}
	\begin{problem}
	\label{opt_problem}
	For a given $\mathbf{c}$ and $\gamma = \frac{1}{\sigma^{2}}$, solve
	\begin{align*}
	q_{1}^{*},q_{2}^{*},\ldots q_{N}^{*}=\arg\underset{q_{1},q_{2},\ldots q_{N}}{\text{min}}\ p_{d} \\
	\text{subject to} \ q_{k} \geq 1, q_{k}\in \mathbb{Z_{+}} \ \forall k \in [N], \sum_{i=1}^{N}q_{i} = \ q_{sum}.  
	\end{align*}
	\end{problem}
	\end{mdframed}
	\vspace{0.5cm}
	

\section{Analysis on the Optimal ARQ Distribution using CC-HARQ}
\label{sec:analysis}

The PDP expression in \eqref{eq:pdp_expression} is dependent on the Marcum-Q function, and tackling the Marcum-Q function analytically is challenging because it contains a modified Bessel function of first kind \cite{Marcum}. Therefore, in this section, we first present an approximation of the Marcum-Q function at a high SNR and then use it to provide necessary and sufficient conditions on the optimal ARQ distribution when using the approximations of the Marcum-Q function. 

\subsection{Approximation on the Marcum-Q function}
\label{approximation_1}
\begin{theorem}
	For a given $N$-hop network and at high SNR regime, we can approximate the Marcum-Q function as $Q_{1}(a_{k},b_{k}) \approx  \tilde{Q}_{1}(a_{k},b_{k}) \triangleq 1 - \frac{b_{k}^{2}}{2} e^{\frac{-a_{k}^{2}}{2}}$ for all $k \in [N]$.
\end{theorem}
\begin{IEEEproof} 
	In the context of the $N$-hop network discussed in the previous section, the Marcum-Q function of first-order associated with the $k$-th hop is given by 
	\begin{eqnarray*}
		e^{-\frac{a_{k}^{2}}{2}}e^{-\frac{b_{k}^{2}}{2}}\sum_{i=0}^{\infty}\bigg(\frac{a_{k}^{2}}{2}\bigg)^{i} \bigg(\sum_{m=0}^{\infty} \bigg(\frac{a_{k}b_{k}}{2}\bigg)^{2m}  \frac{1}{m!(m+i)!}\bigg),  
	\end{eqnarray*}
	where $a_{k}= \sqrt{\frac{2c_{k}}{(1-c_{k})}}$, $b_{k}= \sqrt{\frac{2(2^{R}-1)}{\gamma(1-c_{k})}}$ such that $\gamma = \frac{1}{\sigma^{2}}$. We highlight that $a_{k}$ depends on the LOS component and $b_{k}$ is dependent on SNR. It can be observed that at high SNR, $b_{k}$ is very small, and henceforth, we use the small values of $b_{k}$ to derive the approximation. On expanding $e^{-\frac{b_{k}^{2}}{2}}$ using Taylor series, and neglecting the higher power terms, we can approximate $e^{\frac{-b_{k}^{2}}{2}}$ by $(1-\frac{b_{k}^{2}}{2})$. Similarly, by expanding the internal summation starting with index $m$, we can neglect the higher power terms of $b_{k}$ and approximate the summation terms as $$\sum_{m=0}^{\infty}\bigg(\frac{a_{k}b_{k}}{2}\bigg)^{2m} \frac{1}{m!\Gamma(m+i+1)} \approx \frac{1}{i!}+\frac{1}{(1+i)!}\bigg(\frac{a_{k}^{2}}{2}\bigg)b_{k}^{2}.$$ Therefore, we can rewrite $Q_{1}(a_{k},b_{k})$ as   
	\begin{eqnarray*}
		Q_{1}(a_{k},b_{k})& \approx &e^{-\frac{a_{k}^{2}}{2}}\bigg(1- \frac{b_{k}^{2}}{2} \bigg)\sum_{i=0}^{\infty}\bigg(\frac{a_{k}^{2}}{2}\bigg)^{i} \bigg( \frac{1}{i!}+  \frac{1}{(1+i)!} \\ & & \bigg(\frac{a_{k}^{2}}{4} \bigg)b_{k}^{2}\bigg).   
	\end{eqnarray*}
	Now, by using $\sum_{n=0}^{\infty}x^{n}/n! = e^{x}$ and re-arranging the terms in above equation, we can write
	\begin{eqnarray*}
		Q_{1}(a_{k},b_{k}) & \approx & e^{-\frac{a_{k}^{2}}{2}}\bigg(1- \frac{b_{k}^{2}}{2} \bigg)\bigg[e^{\frac{a_{k}^{2}}{2}}  + \frac{b_{k}^{2}}{2} \big(e^{\frac{a_{k}^{2}}{2}} - 1 \big)\bigg].    
	\end{eqnarray*}
	On further solving the above equation and by neglecting the higher power terms of $b_{k}$, we obtain
	\begin{equation}
		\label{eq:approximation_marcum}
		Q_{1}(a_{k},b_{k}) \approx \tilde{Q}_{1}(a_{k},b_{k}) \triangleq 1 - \frac{b_{k}^{2}}{2} e^{\frac{-a_{k}^{2}}{2}}.
	\end{equation}
	This completes the proof.
	\begin{figure}[h!]
		\centering \includegraphics[scale = 0.45]{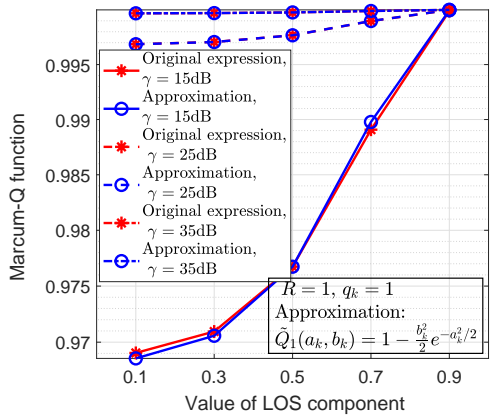}\centering{\caption{  Comparison of original expression of Marcum Q-function with its proposed approximation at different values of SNR.
				\label{fig:Marcum_approx}}}
	\end{figure}
\end{IEEEproof}

To validate the approximation given in \eqref{eq:approximation_marcum} for different values of SNR and the LOS components, we present the simulation results in Fig. \ref{fig:Marcum_approx} where the original expression of Marcum-Q function is compared with its approximated expression given in \eqref{eq:approximation_marcum}. It can be observed from Fig. \ref{fig:Marcum_approx} that the approximation on the Marcum-Q function almost coincides with the original expression. 

By using the approximation in \eqref{eq:approximation_marcum}, we can approximate \eqref{eq:outage_prob_link} as
\begin{equation*}
	\tilde{P}_{k}  = 1- \tilde{Q}_{1}(a_{k},b_{k}) \triangleq \frac{b_{k}^{2}}{2} e^{\frac{-a_{k}^{2}}{2}} = \frac{\phi}{(1-c_{k})} e^{\frac{-a_{k}^{2}}{2}},
\end{equation*}
where $\phi = \frac{2^{R}-1}{\gamma}$, and $\tilde{P}_{k}$ denotes the approximation on $P_{k}$. Similarly, for a given $q_{k}$ retransmissions, we can approximate $P_{kq_{k}}$ as $\tilde{P}_{kq_{k}}$, given by
\begin{equation}
	\label{eq:approximation_P_k}
	\tilde{P}_{kq_{k}} \triangleq \frac{\phi}{q_{k}(1-c_{k})} e^{\frac{-a_{k}^{2}}{2}}.
\end{equation}
Finally, by using \eqref{eq:approximation_P_k}, we can write the approximate version of \eqref{eq:pdp_expression} as
\begin{equation}
	\label{eq:Approx_pdp_expression}
	\tilde{p}_{d} = \tilde{P}_{1q_{1}} + \sum_{k=2}^{N} \tilde{P}_{kq_{k}} \Bigg( \prod_{j=1}^{k-1}  (1-\tilde{P}_{jq_{j}})\Bigg),
\end{equation} 
where $\tilde{p}_{d}$ denotes the approximation on $p_{d}$. Henceforth, using the approximated expression on $p_{d}$ as above, we formulate an  optimization problem in Problem \ref{opt_problem_2} as shown below. Throughout this paper, we refer to the solution of Problem \ref{opt_problem_2}, as the near-optimal ARQ distribution since the objective function in Problem \ref{opt_problem_2} is an approximation on the objective function in Problem \ref{opt_problem}. 
\vspace{0.5cm}
\begin{mdframed}
	\begin{problem}
		\label{opt_problem_2}
		For a given $\mathbf{c}$ and a high SNR $\gamma = \frac{1}{\sigma^{2}}$, solve
		\begin{align*}
			q_{1}^{*},q_{2}^{*},\ldots q_{N}^{*}=\arg\underset{q_{1},q_{2},\ldots q_{N}}{\text{min}}\ \tilde{p}_{d} \\
			\text{subject to} \ q_{k} \geq 1, q_{k}\in \mathbb{Z_{+}} \ \forall k,
			\sum_{i=1}^{N}q_{i} = \ q_{sum}.  
		\end{align*}
	\end{problem}
\end{mdframed}
\vspace{0.5cm}


  
\subsection{Sufficient and Necessary Conditions on the Near-Optimal ARQ Distribution in CC-HARQ Strategy}
\label{sec:neccAndSuffConditions}

Before we obtain the necessary and sufficient conditions on the optimal ARQ distribution, we show that a link with a higher LOS component must not be given more ARQs than the link with a lower LOS component.

\begin{theorem}
	\label{LOShighlow}
	For a given LOS vector $\mathbf{c}$ and a high SNR regime in a CC-HARQ based multi-hop network, the solution to Problem \ref{opt_problem_2} satisfies the property that whenever $c_{i} \geq c_{j}$, we have $q_{i} \leq q_{j} \ \forall~ i,j \ \in [N]$. 
\end{theorem}
\begin{IEEEproof}
	To highlight the location of $c_{i}$ and $c_{j}$, we write $\mathbf{c}$ as  $[c_{1}, c_{2}, \ldots, c_{i}, \ldots, c_{j}, \ldots, c_{N-1}, c_{N}]$ such that $j > i$. Assume that $c_{j} > c_{i}$, and $q_{i}$ and $q_{j}$ denote the number of ARQs allotted to the $i$-th and $j$-th link, respectively. Furthermore, let us suppose that $q_{i} =q_{j} = q$ and we have one ARQ with us. Now, the problem is deciding whether to allocate that additional ARQ to the $i$-th link or the $j$-link that results in lower PDP. Towards solving this problem, we consider an equivalent multi-hop network with LOS vector $\mathbf{c}' = [c_{1}, c_{2}, \ldots, c_{N-1}, \ldots, c_{N}, \ldots, c_{i}, c_{j}]$, wherein $\mathbf{c}'$ is obtained from $\mathbf{c}$ by swapping $c_{i}$ with $c_{N-1}$ and $c_{j}$ with $c_{N}$. By using a similar approach given in \cite[Theorem 1]{our_work_TWC_1}, we can show that the PDP of the multi-hop networks with the LOS vectors $\mathbf{c}$ and $\mathbf{c}'$ are identical. Furthermore, the PDP of the $N$-hop network with LOS vector $\mathbf{c}'$, is written as 
	\begin{eqnarray*}
		\tilde{p}_{d} & = & \tilde{P}_{1q_{1}} + \tilde{P}_{2q_{2}}(1-\tilde{P}_{1q_{1}}) + \ldots \\
		& & + \left(\tilde{P}_{iq_{i}} + \tilde{P}_{jq_{j}}(1 -\tilde{P}_{iq_{i}})\right)\prod_{k \in [N] \setminus \{i, j\}} (1-\tilde{P}_{kq_{k}}). 
	\end{eqnarray*} 
	Note that $\tilde{P}_{iq_{i}}$ and $\tilde{P}_{jq_{j}}$ appear only in the last term of the above expression. Since the question of allocating the additional ARQ is dependent only on the expression $\tilde{P}_{iq_{i}} + \tilde{P}_{jq_{j}}(1 - \tilde{P}_{iq_{i}})$, we henceforth do not use the entire expression for PDP. Additionally, since $q_{i} = q_{j} = q$, 
	we obtain one of the following expressions when allocating the additional ARQ,  
	\begin{eqnarray*}
		A & =& \tilde{P}_{i(q+1)}+\tilde{P}_{jq}(1-\tilde{P}_{i(q+1)}),\\
		B & =& \tilde{P}_{iq} + \tilde{P}_{j(q+1)}(1-\tilde{P}_{iq}).
	\end{eqnarray*}
	Since $c_{i} < c_{j}$, we know that $\tilde{P}_{iq} > \tilde{P}_{jq}$ for $q=1$. To prove the statement of the theorem, we use the approximation given in \eqref{eq:approximation_P_k} to show that $A < B$. Therefore, using \eqref{eq:approximation_P_k}, we write 
	\begin{eqnarray*}
		\begin{aligned}
			A = \frac{\phi e^{\frac{-a_{i}^{2}}{2}}}{(q+1)(1-c_{i})}  + \frac{\phi e^{\frac{-a_{j}^{2}}{2}}}{q(1-c_{j})}\bigg(1-  \frac{\phi e^{\frac{-a_{i}^{2}}{2}}}{(q+1)(1-c_{i})}\bigg),\\
			B  = \frac{\phi e^{\frac{-a_{i}^{2}}{2}}}{q(1-c_{i})}  + \frac{\phi e^{\frac{-a_{j}^{2}}{2}}}{(q+1)(1-c_{j})}\bigg(1-  \frac{\phi e^{\frac{-a_{i}^{2}}{2}}}{q(1-c_{i})}\bigg).	
		\end{aligned}
	\end{eqnarray*}
		On solving the difference $A-B$, we get 
\begin{eqnarray*}
  \phi\bigg( \dfrac{e^{-a_{j}^{2}/2}-e^{-a_{i}^{2}/2}+ c_{j}e^{-a_{i}^{2}/2}-c_{i}e^{-a_{j}^{2}/2}}{{(1-c_{i})(1-c_{j})}}\bigg ) \underbrace{\bigg(\dfrac{1}{q}-\dfrac{1}{q+1}\bigg)}. 
\end{eqnarray*}
In the above equation, note that the second product term in the bracket is positive. Therefore, to prove that $A-B<0$, first, we show that $e^{-a_{j}^{2}/2}-e^{-a_{i}^{2}/2}+ c_{j}e^{-a_{i}^{2}/2}-c_{i}e^{-a_{j}^{2}/2} \leq 0$ for any $i,j \in [N]$. To proceed further, we start by assuming that the above inequality is true, and by rewriting the above equation, we get $e^{-a_{j}^{2}/2}(1-c_{i})  \leq e^{-a_{i}^{2}/2}(1-c_{j})$ which in turns equal to $\dfrac{(1-c_{i})}{(1-c_{j})} \leq \dfrac{e^{-a_{i}^{2}/2}}{e^{-a_{j}^{2}/2}}$. Now, by taking the logarithm on both sides and on expanding both $a_{i}$ and $a_{j}$, we obtain the inequality $\log\bigg(\dfrac{1-c_{i}}{1-c_{j}}\bigg) \leq \dfrac{c_{j}-c_{i}}{(1-c_{i})(1-c_{j})}$. Let $x =\dfrac{1-c_{i}}{1-c_{j}} $, and therefore, $x-1 = \dfrac{1-c_{i}}{1-c_{j}} - 1 = \dfrac{c_{j}-c_{i}}{(1-c_{j})}$. Furthermore, by using $x$, we can rewrite the above inequality as $\log x \leq \dfrac{x-1}{(1-c_{i})}$, where $\dfrac{1}{1-c_{i}} \geq 1$ because $c_{i} \leq 1$. Moreover, by using the standard inequality i.e. $\log x \leq (x-1)$, we can prove that the inequality $\log x \leq \dfrac{x-1}{(1-c_{i})}$ is true for any $i, j \in [N]$. Hence, this completes the proof that $A-B<0$.  
\end{IEEEproof}
Henceforth, we use the set $\mathbb{S} = \{\mathbf{q} \in \mathbb{Z}_{+}^{N} \ | \ \sum_{j=1}^{N} q_{j} = q_{sum} \ \& \ q_{j} \geq 1 \ \forall~ j \} $ to define the search space for the optimal ARQ distribution. Furthermore, for a given $\mathbf{q} \in \mathbb{S}$, its neighbors are defined as below.
\begin{definition}
\label{def:nn}
The set of neighbors for a given $\mathbf{q} \in \mathbb{S}$ is defined as $\mathcal{H}(\mathbf{q}) = \{\bar{\mathbf{q}} \in \mathbb{S} ~|~ d(\mathbf{q},\bar{\mathbf{q}}) = 2\},$ where $d(\mathbf{q},\bar{\mathbf{q}})$ denotes the Hamming distance between $\mathbf{q}$ and $\bar{\mathbf{q}}$.
\end{definition}
In the following definition, we present a local minima of the space $\mathbb{S}$ by evaluating the PDP of the CC-HARQ based multi-hop network over the vectors in $\mathbb{S}$.  
\begin{definition}
\label{def:lm}  
To be a local minima of $\mathbb{S}$, an ARQ distribution $\mathbf{q}^{*} \in \mathbb{S}$ must satisfy the condition $p_{d}(\mathbf{q}^{*}) \leq p_{d}(\mathbf{q}), \mbox{ for every } \mathbf{q} \in \mathcal{H}(\mathbf{q}^{*})$, such that $p_{d}(\mathbf{q}^{*})$ and $p_{d}(\mathbf{q})$ represent the PDP evaluated at the distributions $\mathbf{q}^{*}$ and $\mathbf{q}$, respectively.   
\end{definition}
Using the above definition, we derive a set of necessary and sufficient conditions on the local minima in the following theorem.
\begin{theorem}
\label{NeccSuff}
For a given $N$-hop network with LOS vector $\mathbf{c}$, the ARQ distribution $\mathbf{q}^{*} = [q^{*}_{1}, q^{*}_{2}, \ldots, q^{*}_{N}]$ is said to be a local minima if and only if $q^{*}_{i}$ and $q^{*}_{j}$ for $i \neq j$ satisfy the following bounds
\begin{eqnarray}
\label{LBUB_1}
q^{*2}_{j}K_{i}- q^{*}_{j}(K_{i}+K_{j}K_{i}) -C_{1} & \leq & 0  ,\\
\label{LBUB_2}
q^{*2}_{j}K_{i}+ q^{*}_{j}(K_{i}-K_{j}K_{i})-C_{2} & \geq & 0,
\end{eqnarray} 
\noindent where $C_{1} = -K_{j}K_{i} + q^{*2}_{i}K_{j} +  q^{*}_{i}(K_{j}-K_{j}K_{i}) $, $C_{2} = K_{j}K_{i} + q^{*2}_{i}K_{j}-  q^{*}_{i}(K_{j}+K_{j}K_{i}) $ with $K_{t}$ for $t \in \{i,j\}$ given by
\begin{eqnarray}
\label{eq:Kt}
K_{t} =\dfrac{\phi}{1-c_{t}}e^{\big(\dfrac{-c_{t}}{1-c_{t}}\big)}.
\end{eqnarray}
\end{theorem}	
\begin{IEEEproof}
According to Definition \ref{def:nn}, it can be observed that a neighbor of $\mathbf{q}^{*}$ in the search space $\mathbb{S}$ differs in two positions with respect to $\mathbf{q}^{*}$. Let these neighbors are of the form $\hat{\mathbf{q}}_{+} = [q^{*}_{1}, q^{*}_{2}, \ldots, q^{*}_{i} + 1, \ldots, q^{*}_{j} -1, \ldots, q^{*}_{N}]$ and $\hat{\mathbf{q}}_{-} = [q^{*}_{1}, q^{*}_{2}, \ldots, q^{*}_{i} - 1, \ldots, q^{*}_{j} + 1, \ldots, q^{*}_{N}]$ that differs in two positions at $i$ and $j$ provided $q^{*}_{i} - 1 \geq 1$ and $q^{*}_{j} - 1 \geq 1$. Because of the type of search space and the expression of PDP, we invoke the results from \cite[Theorem 1]{our_work_TWC_1} that the PDP remains identical after swapping intermediate links. Therefore, instead of considering the multi-hop network with LOS vector $\mathbf{c} = [c_{1}, c_{2}, \ldots, c_{i}, \ldots, c_{j}, \ldots, c_{N-1}, c_{N}]$, we consider its permuted version with the LOS vector $\mathbf{c} = [c_{1}, c_{2}, \ldots, c_{N-1}, \ldots, c_{N}, \ldots, c_{i}, c_{j}]$, wherein the $i$-th link is swapped with $(N-1)$-th link, and the $j$-th link is swapped with $N$-th link. Consequently, the local minima and its two neighbors are respectively of the form $\mathbf{q}^{*} = [q^{*}_{1}, q^{*}_{2}, \ldots, q^{*}_{N-1}, \ldots, q^{*}_{N}, \ldots, q^{*}_{i}, q^{*}_{j}]$, $\hat{\mathbf{q}}_{+} = [q^{*}_{1}, q^{*}_{2}, \ldots, q^{*}_{N-1}, \ldots, q^{*}_{N}, \ldots, q^{*}_{i} + 1, q^{*}_{j} - 1]$ and $\hat{\mathbf{q}}_{-} = [q^{*}_{1}, q^{*}_{2}, \ldots, q^{*}_{N-1}, \ldots, q^{*}_{N}, \ldots, q^{*}_{i} - 1, q^{*}_{j} + 1]$. By using the definition of local minima, we have the inequalities
\begin{eqnarray}
\label{eqr1}
\tilde{p}_{d}(\mathbf{q}^{*}) \leq \tilde{p}_{d}(\hat{\mathbf{q}}_{+}), \mbox{ and } \tilde{p}_{d}(\mathbf{q}^{*}) \leq \tilde{p}_{d}(\hat{\mathbf{q}}_{-}),
\end{eqnarray}
where $\tilde{p}_{d}(\mathbf{q}^{*})$, $\tilde{p}_{d}(\hat{\mathbf{q}}_{+})$ and $\tilde{p}_{d}(\hat{\mathbf{q}}_{-})$ represent the PDP evaluated at the distributions $\mathbf{q}^{*}$, $\hat{\mathbf{q}}_{+}$, and $\hat{\mathbf{q}}_{-}$, respectively. Because of the fact that $\hat{\mathbf{q}}_{+}$ and $\hat{\mathbf{q}}_{-}$ differ only in the last two positions and the structure of the PDP, it is possible to demonstrate that the inequalities in $\eqref{eqr1}$ are equivalent to 
\begin{eqnarray}
\label{express1}
\tilde{P}_{iq^{*}_{i}}+ \tilde{P}_{jq^{*}_{j}}\left(1-\tilde{P}_{iq^{*}_{i}}\right) \leq  \tilde{P}_{i(q^{*}_{i}+1)} + \tilde{P}_{j(q^{*}_{j}-1)} (1- \nonumber \\ \tilde{P}_{i(q^{*}_{i}+1)}),\\
\label{express2}
\tilde{P}_{iq^{*}_{i}}+\tilde{P}_{jq^{*}_{j}}\left(1-\tilde{P}_{iq^{*}_{j}}\right) \leq \tilde{P}_{i(q^{*}_{i}-1)} + \tilde{P}_{j(q^{*}_{j}+1)}(1- \nonumber \\ \tilde{P}_{i(q^{*}_{i}-1)}),	
\end{eqnarray} respectively. 
First, let us proceed with \eqref{express1} to derive a necessary and sufficient conditions on $q^{*}_{i}$ and $q^{*}_{j}$. On expanding $\tilde{P}_{iq}$ and $\tilde{P}_{jq}$, and by using the approximation on outage probability given in \eqref{eq:approximation_P_k}, the inequality in \eqref{express1} can be rewritten as 
\begin{equation*}
\dfrac{K_{i}}{q_{i}^{*}} + 	\dfrac{K_{j}}{q_{j}^{*}}\bigg(1-\dfrac{K_{i}}{q_{i}^{*}}\bigg) \leq \dfrac{K_{i}}{q_{i}^{*}+1} + 	\dfrac{K_{j}}{q_{j}^{*}-1}\bigg(1-\dfrac{K_{i}}{q_{i}^{*}+1}\bigg),
		\end{equation*}
	where $K_{i}$ and $K_{j}$ can be obtained from \eqref{eq:Kt}. On solving the above equation, we obtain
		\begin{equation*}
			\dfrac{K_{i}}{q_{i}^{*}}-\dfrac{K_{i}}{q_{i}^{*}+1} + \dfrac{K_{j}}{q_{j}^{*}}-\dfrac{K_{j}}{q_{j}^{*}-1}	\leq \dfrac{K_{i}K_{j}}{q_{i}^{*}q_{j}^{*}} - \dfrac{K_{i}K_{j}}{(q_{i}^{*}+1)(q_{j}^{*}-1)}.
		\end{equation*}
		After further modifications, we can rewrite the above equation as 
		\begin{eqnarray}
			\label{eq:true_inequality_1}
			q_{j}^{*2}K_{i}-q_{j}^{*}(K_{i}+K_{i}K_{j})  \leq  - K_{i}K_{j} +  q_{i}^{*2}K_{j} +  q_{i}^{*} \nonumber \\ (K_{j}- K_{i}K_{j}).
		\end{eqnarray}
		In the above equation, we can replace $ (-K_{i}K_{j}+q_{i}^{*2}K_{j} +q_{i}^{*}(K_{j}- K_{i}K_{j}))$ by $C_{1}$, and by rearranging the terms, we get \eqref{LBUB_1}. This completes the proof for the first necessary condition. The second necessary condition for \eqref{express2} can be proved along the similar lines of the above proof to obtain \eqref{LBUB_2}. It can be observed that the two conditions of this theorem are also sufficient because the bounds are obtained by rearranging the terms in the condition on local minima.
\end{IEEEproof}

	\section{Low-complexity List-Decoding Algorithm}	
	\label{sec:proposed_method}

Using Theorem \ref{NeccSuff}, we are ready to synthesize a low-complexity algorithm to solve Problem \ref{opt_problem_2}.
	\begin{proposition}
		\label{prop_1}
		If the ARQ distribution $\mathbf{q}$ is chosen such that $\dfrac{q_{j}^{*}}{q_{i}^{*}} = \sqrt{\dfrac{K_{j}}{K_{i}}}$, for $i \neq j$, then $\mathbf{q}$ is a local minima of the search space at a high SNR. 
	\end{proposition}
	\begin{IEEEproof}
 	 At a high SNR, $\phi$ is very small, and therefore, the product $K_{i}K_{j}$ is negligible. Therefore, we can rewrite \eqref{LBUB_1} and \eqref{LBUB_2} as     
	\begin{eqnarray*}
		q_{j}^{*2}K_{i}-q_{j}^{*}K_{i} &\leq &q_{i}^{*2}K_{j} + q_{i}^{*}K_{j}, \\
		q_{j}^{*2}K_{i}+q_{j}^{*}K_{i} &\geq & q_{i}^{*2}K_{j} - q_{i}^{*}K_{j},
	\end{eqnarray*}	
	respectively. On rearranging the above equations, we get
	\begin{eqnarray*}
		q_{j}^{*2}K_{i}-q_{i}^{*2}K_{j} &\leq &q_{j}^{*}K_{i} + q_{i}^{*}K_{j}, \\
		q_{j}^{*2}K_{i}-q_{i}^{*2}K_{j} &\geq & -q_{j}^{*}K_{i} - q_{i}^{*}K_{j},
	\end{eqnarray*}	
	respectively. Now, by equating $q_{j}^{*}K_{i} + q_{i}^{*}K_{j} = \epsilon_{i,j}$, the above equations become
	\begin{eqnarray}
		\label{eq:approx_inequality_1}
		q_{j}^{*2}K_{i}-q_{i}^{*2}K_{j} &\leq & \epsilon_{i,j}, \\
		\label{eq:approx_inequality_2}
		q_{j}^{*2}K_{i}-q_{i}^{*2}K_{j} &\geq & - \epsilon_{i,j},
	\end{eqnarray}	
	where $\epsilon_{i,j}$ is a positive number dependent on $K_{i}, K_{j}, q_{i}^{*}, q_{j}^{*}$. If we choose $q_{i}^{*}$ and $q_{j}^{*}$ such that $q_{j}^{*2}K_{i}-q_{i}^{*2}K_{j} = 0$, for $i \neq j$, then it ensures that the inequalities in \eqref{eq:approx_inequality_1} and \eqref{eq:approx_inequality_2} are trivially satisfied. Thus, choosing 
	\begin{equation}
		\label{eq:succ_bound}
		\dfrac{q_{j}^{*}}{q_{i}^{*}} = \sqrt{\dfrac{K_{j}}{K_{i}}},
	\end{equation}
	 for every pair $i, j \in [N]$ such that $i \neq j$, satisfies the sufficient conditions given in \eqref{eq:approx_inequality_1} and \eqref{eq:approx_inequality_2} at high SNR values. This completes the proof.	\end{IEEEproof}
	
	Based on the results in Proposition \ref{prop_1}, we formulate Problem \ref{relaxed_problem}, as given below, as a means of solving Problem \ref{opt_problem_2} at a high SNR. 
\vspace{0.5cm}
	\begin{mdframed}
	\begin{problem}
	\label{relaxed_problem}
	For a given $\{K_{1}, K_{2}, \ldots, K_{N}\}$ and $q_{sum}$, find $q_{1}^{*},q_{2}^{*},\ldots q_{N}^{*}$ such that $$\frac{q_{j}^{*}}{q_{i}^{*}} = \sqrt{\dfrac{K_{j}}{K_{i}}}, ~\forall~ i, j \in [N] \ \mbox{where} \ i \neq j,$$ $q_{i}^{*} \geq 1, q_{i}^{*}\in \mathbb{Z_{+}}, ~\forall i, \ \sum_{i=1}^{N}q_{i}^{*} = \ q_{sum}.$  
	\end{problem}
	\end{mdframed}
	\vspace{0.5cm}	
	However, from Problem \ref{relaxed_problem}, it is observed that a solution is not guaranteed because the ratio $\sqrt{\dfrac{K_{j}}{K_{i}}}$, which is computed based on the LOS components and the SNR, need not be in $\mathbb{Q}$. Therefore, we first propose a method to solve Problem \ref{relaxed_problem} without the integer constraints, i.e., to find an ARQ distribution $\mathbf{q} \in \mathbb{R}^{N}$. This type of problem can be solved by using the system of linear equations of the form $\mathbf{R} \mathbf{q} = \mathbf{s}$ to obtain $\mathbf{q}_{real} = \mathbf{R}^{-1}\mathbf{s},$ where $\mathbf{q} = [q_{1}, q_{2}, \ldots, q_{N}]^{T}$, $\mathbf{s} = [0, 0, \ldots,0, q_{sum}]^{T}$ and $\mathbf{R} \in \mathbb{R}^{N \times N}$ such that $\mathbf{R}(j, j) = 1$ for $1 \leq j \leq N$, $\mathbf{R}(N, j) = 1$, for $1 \leq j \leq N$, $\mathbf{R}(j, j+1) = -r_{j, j+1}$, for $r_{j,j+1} \triangleq \sqrt{\dfrac{K_{j+1}}{K_{j}}}$ and $1 \leq j \leq N-1$, and the rest of the entries of $\mathbf{R}$ are zeros. Subsequently, using the solution $\mathbf{q}_{real}$ in $\mathbb{R}^{N}$, we propose to obtain a solution in the true search space $\mathbb{S}$ by generating a list as given in Algorithm \ref{List algoritm} \cite{wiopt}. Finally, once the list is generated, the ARQ distribution that minimizes the PDP would be the solution of the proposed algorithm. 
	\begin{algorithm}
		\caption{\label{List algoritm}List Creation Based Algorithm}
		\label{LIST_algorithm}
		\begin{algorithmic}[1]
			\Require $\mathbf{R}$, $\mathbf{s}$, $q_{sum}$, $\mathbf{c} = [c_{1}, c_{2}, \ldots, c_{N}]$
			\Ensure $\mathcal{L} \subset \mathbb{S}$ - List of ARQ distributions in search space $\mathbb{S}$.
			\State Compute $\mathbf{q}_{real} = \mathbf{R}^{-1}\mathbf{s}$, $\tilde{\mathbf{q}} = \lceil \mathbf{q}_{real} \rceil$.
			\For {$i = 1:N$} 
			\If {$\tilde{q}_{i} = 0$} 
			\State $\tilde{q}_{i} = \tilde{q}_{i} + 1$			
			\EndIf
			\EndFor
			\State Compute $E = \left(\sum_{i = 1}^{N} \tilde{q}_{i}\right) - q_{sum}$
			\State $\mathcal{L} = \{\mathbf{q} \in \mathbb{S} ~|~ d(\mathbf{q}, \tilde{\mathbf{q}}) = E, q_{j} \ngtr q_{i} \text{ for }  c_{i} < c_{j}\}$. 
		\end{algorithmic}
	\end{algorithm}
\begin{figure*}[t]
\begin{center}
\includegraphics[width=18.2cm, height=6.0cm]{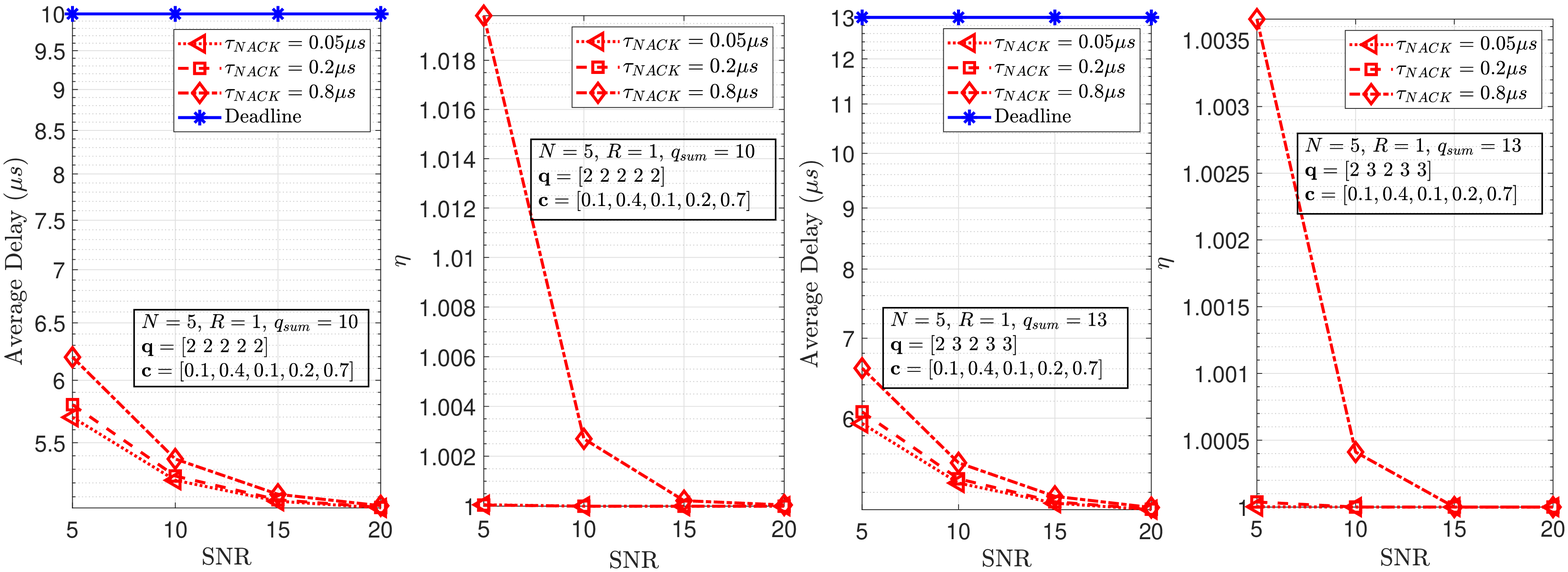}
\vspace{-0.5cm}	
\centering{\caption{Illustration on the average delay on packets and deadline violation parameter ($\eta$) for several values of $\tau_{NACK} = \{0.05, 0.2, 0.8\}$ microseconds while implementing CC-HARQ strategy.
\label{fig:average_delay}}}
\end{center}
\end{figure*}
	\begin{figure}[h!]
	\centering \includegraphics[scale = 0.41]{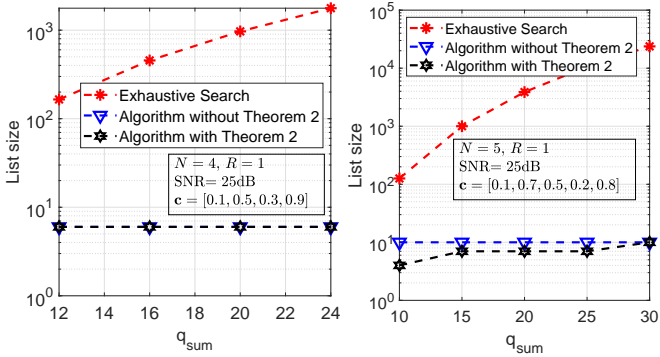}
	\vspace{-0.5cm}	
	\centering{\caption{Comparison of list size between exhaustive search and the proposed low-complexity algorithm with and without Theorem \ref{LOShighlow} .
	\label{fig:List_size}}}
\end{figure}	
\begin{figure*}[h]
	\begin{center}
		\includegraphics[scale = 0.57]{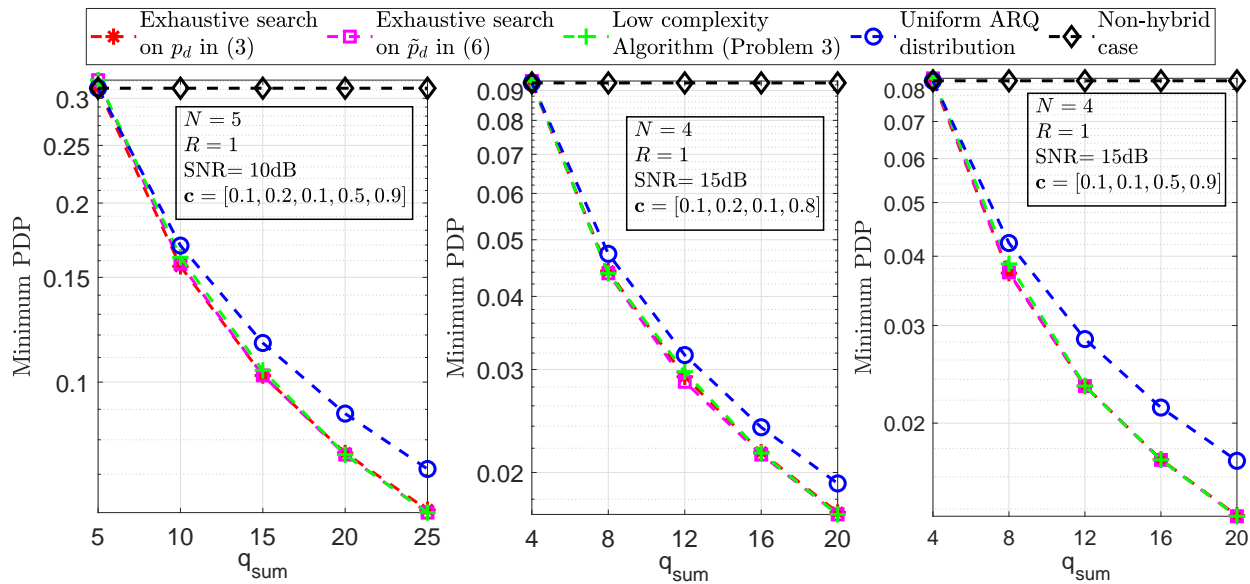}{\caption{PDP comparison between (i) an exhaustive search using original PDP expression in CC-HARQ strategy, (ii) an exhaustive search using approximated PDP expression in CC-HARQ strategy, (iii) the proposed low-complexity algorithm, (iv) uniform ARQ distribution in CC-HARQ based strategy, and (v) ARQ distribution in non-HARQ strategy \cite{our_work_TWC_1}.     
				\label{fig:pdp_comparison}}}
	\end{center}
\end{figure*}	

	\section{Simulation Results}
	\label{sec:Sims}
	
In the first part of this section, we present simulation results on the delay profiles of the packets under the CC-HARQ protocol. We show that when the ACK/NACK delay overheads are sufficiently small as compared to the payload, there is a high probability with which the packets arrive at the destination within the given deadline. To obtain the results, first, we take $q_{sum} = \floor{\frac{\tau_{total}}{\tau_{p}  + \tau_{d}}}$, by assuming $\tau_{NACK} = 0$ and $\tau_{p} + \tau_{d} = 1$ microsecond where $\tau_{total}, \tau_{NACK}, \tau_{p} \ \text{and} \ \tau_{d}$ (with $\mu s$ unit) are defined in Section \ref{sec:intro}. Then, we consider non-zero values of $\tau_{NACK}= \{0.05, 0.2, 0.8\} \mu s $ during the packet flow, and present the results for the following metrics: (i) the number of packets that did not reach the destination due to lack of ARQs present at the intermediate nodes (defined by $P_{drop}$), (ii) the number of packets that reach the destination after the given deadline (defined by $P_{deadline}$), and (iii) the average delay on the packets. The plots are as shown in Fig. \ref{fig:average_delay} for a given $N$ and a LOS vector $\textbf{c}$ with different $q_{sum}$ and various SNR values. It can be observed that when $\tau_{NACK}$ is very small, specifically at values $\tau_{NACK}=\{0.05, 0.2\}$ microseconds, the average delay is sufficiently lower than the $q_{sum}$ (the deadline). Also, the deadline violation parameter $\eta = \frac{P_{drop} + P_{deadline}}{P_{drop}}$ is nearly or equal (in the case of $\tau_{NACK}=0.05 \mu s$) to one. However, when $\tau_{NACK}=0.8 \mu s$, the average delay slightly moves towards the deadline and also, $\eta$ exceeds one, showing the deadline violation of the packets.      
 
In the rest of this section, we present the simulation results to validate our theoretical analysis and to showcase the benefits of using the CC-HARQ protocol over a non-HARQ based multi-hop model \cite{our_work_TWC_1}. On the one hand, the computational complexity for solving Problem \ref{opt_problem} and Problem \ref{opt_problem_2} using an exhaustive search is ${q_{sum}-1}\choose{N-1}$. On the other hand, the computational complexity of our method is determined by the complexity of computing the inverse of a matrix and that of Algorithm \ref{LIST_algorithm}. Furthermore, while the complexity for computing the inverse is $O(N^{3})$, the number of computations for generating the list is bounded by ${N}\choose{E}$, where $E$ is the leftover number of ARQs after the ceiling operation. To demonstrate the benefits of using our low-complexity method, we plot the curves for the list size of an exhaustive search and the proposed algorithm in Fig. \ref{fig:List_size} for several values of $q_{sum}$ with $N = 4$ and $N = 5$. Specifically, we plot the list size both with and without incorporating the results of Theorem \ref{LOShighlow}. When using Theorem \ref{LOShighlow}, by subtracting one ARQ from all possible $\binom{N}{E}$ positions from $\tilde{\mathbf{q}}$, we discard those ARQ distributions which follow the rule $q_{i} > q_{j}$ whenever $c_{i} > c_{j}$. As a result, we see that the list size shortens after incorporating the rule of Theorem \ref{LOShighlow} for $N = 5$. Furthermore, based on the simulation results, we observe that the ARQ distribution, which minimizes the PDP from the list $\mathcal{L}$ matches the result of exhaustive search, thereby confirming that our list contains the optimal ARQ distribution. Although, we used high SNR results of Proposition \ref{prop_1} to synthesize the list-decoding method, we observe that the size of the list reduces significantly at low and medium-range of SNR values. 

Finally, to showcase the benefits of using CC-HARQ in a multi-hop network, we plot the minimum PDP against several values of $q_{sum}$ in Fig. \ref{fig:pdp_comparison} by using (i) an exhaustive search on $p_{d}$ (as given in \eqref{eq:pdp_expression}), (ii) an exhaustive search on $\tilde{p}_{d}$ (as given in \eqref{eq:Approx_pdp_expression}), (iii) the proposed low-complexity algorithm, (iv) uniform ARQ distribution in CC-HARQ based strategy, and (v) non-HARQ strategy. There are mainly two observations from the simulation results in Fig. \ref{fig:pdp_comparison}: (a) the PDP improves in the case of optimal ARQ distribution (with the exhaustive search on both $p_{d}$ and $\tilde{p}_{d}$) over uniform ARQ distribution and non-hybrid based ARQ distribution, and (b) our proposed approximation $\tilde{p}_{d}$ produces reasonably accurate results.

\section{Discussion}
\label{sec:conclusions}

In this work, we have presented a novel CC-HARQ framework to achieve high reliability with bounded-delay constraints in a slow-fading multi-hop network. In particular, by using the CC-HARQ protocol at the intermediate relay, we have posed the problem of distributing a given number of ARQs such that the PDP is minimized. We have shown that the problem is non-tractable because the underlying PDP expression contains the Marcum-Q function. Towards solving the problem, (i) we have provided a novel approximation on the Marcum-Q function in the high SNR regime and have subsequently used the approximation by formulating a similar optimization problem, (ii) we have derived a set of necessary and sufficient conditions on the near-optimal ARQ distribution by using the approximated version of the optimization problem, and finally, (iii) we have proposed a low-complexity list-based algorithm to solve the optimization problem that yields near-optimal ARQ distribution. We have validated our approximation and the efficacy of our algorithm through extensive simulation results. 
\end{document}